\documentstyle[12pt,prd,aps]{revtex}
\tightenlines
\input epsf
\begin{document}
\preprint{UAB-FT-438,OU-TAP 76}
\title{The density parameter and the Anthropic Principle}
\author{Jaume Garriga} \address{IFAE, Departament de Fisica,
Universitat Autonoma de Barcelona,\protect\\ 08193 Bellaterra
(Barcelona), Spain} 
\author{Takahiro Tanaka} \address{Department of Earth and Space Science,
Graduate School of Science, Osaka University, Toyonaka 560-0043, Japan}
\author{Alexander Vilenkin} \address{Institute of Cosmology, Department of
Physics and Astronomy,\protect\\ Tufts University, Medford, MA
02155, USA} 
\maketitle

\begin{abstract}
In the context of open inflation,
we calculate the probability distribution for the density parameter
$\Omega$.
A large class of two field models of open inflation do not
lead to infinite open universes, but to an ensemble of inflating islands of 
finite size, or ``quasi-open'' universes, where the density parameter 
takes a range of values. Assuming we
are typical observers, the models make definite predictions for
the value $\Omega$ we are most likely to observe. 
When compared with observations, these predictions can be used to constrain 
the parameters of the models. We also argue that
obsevers should not be surprised to find themselves living at the
time when curvature is about to dominate.
\end{abstract}
\pacs{1}

\section{Introduction}

Anthropic considerations have often been used in order to
justify the ``naturalness'' of 
the values taken by certain constants of Nature \cite{anthropic}. 
In these approaches,
it is assumed that the ``constants'' are really random variables whose
range and ``a priori'' probabilities are determined by the laws of Physics. 
Knowledge of these ``a priori'' probabilities is certainly useful,
but not sufficient to determine the probability
for an observer to measure given values of the constants.
For instance, some values which are in the ``a priori''
allowed range may be incompatible with the very existence of observers,
and in this case they will never be measured.
The relevant question is then how to assign a weight to this 
selection effect.  

A natural framework where these ideas can be applied is inflation.
There, the false-vacuum energy of the scalar field which drives the 
inflationary phase can thermalize in different local minima of its
potential, and each local minimum may have a different set of values for
the constants of Nature. Also, there may be different 
routes from false vacuum to a given minimum. In this case all
thermalized regions will have the same low energy Physics constants,
but each route will yield a hot universe with different large scale 
properties. Here, we shall be concerned with this possibility, where the 
fundamental constants (such as the gauge couplings or the cosmological 
constant) are
fixed, but other cosmological parameters such as the density parameter
or the amplitude of cosmological perturbations are random variables
whose distribution is dynamically determined.

In this context, the most reasonable -and predictive- version of 
the anthropic principle seems to be the principle of mediocrity 
\cite{medi,gott}, 
according to which we are typical observers who shall observe 
what the vast majority of observers would. Thus, the measure of probability 
for a given set of constants is simply proportional to the total 
number of civilizations 
emerging with those values of the constants. In this paper we shall use
this principle in order to calculate the probability distribution
for the density parameter $\Omega$.

Standard inflationary models predict $\Omega=1$ with ``certainty''. 
What this means is that these models can explain the observed 
homogeneity and isotropy of the universe only if the universe is flat.
However, a class of ``open inflation'' models which 
lead to $\Omega<1$ have received some attention in recent years
\cite{open,BGT,LM}. 
In these models, inflation proceeds in two steps.
One starts with a scalar field $\sigma$ trapped in a 
metastable minimum of its potential $V(\sigma)$. The false vacuum energy
drives an initial period of exponential expansion, and                  
decays through quantum nucleation of highly symmetric bubbles of true vacuum.
The interior of these bubbles has the geometry of an open 
Friedmann-Robertson-Walker universe. This accounts for the 
observed homogeneity and isotropy of the universe. In order to solve the 
flatness problem a second stage of slow roll inflation inside 
the bubble is necessary.

In models with a single scalar field $\sigma$, all bubbles have the
same value of $\Omega$ which is determined by the number of e-foldings
in the second period of inflation.  
The potential $V(\sigma)$ in such
models is assumed to have a rather special form, with a sharp barrier
next to a flat slow-roll region, which requires a
substantial amount of fine-tuning. Additional tunning is needed to
arrange the desired value of $\Omega$.
A more natural class of models
includes two 
fields, $\sigma$ and $\phi$, with $\sigma$ doing the tunneling and
$\phi$ the slow roll \cite{LM}.  The simplest example is
\begin{equation}
\label{coupled}
V(\sigma,\phi)=V_t(\sigma) + {g \over 2} \sigma^2 \phi^2,
\end{equation}
where $V_0(\sigma)$ has a metastable false vacuum at $\sigma=0$.
After $\sigma$ tunnels to its true minimum $\sigma=v$, 
the field $\phi$ would
drive a second period of slow roll inflation inside the bubble. 
Depending on the 
value of $\phi$ at the time of nucleation, the number of e-foldings
of the second stage of inflation would be different. 

Initially, it was believed \cite{LM} that
models such as (\ref{coupled}) would yield an ensemble of infinite open 
universes, one inside each nucleated bubble, and each one with a different 
value of the density parameter.
However, it has been recently realized \cite{GGM} that this picture is
oversimplified.  The two field models which allow 
for variable $\Omega$
do not actually lead to infinite open 
universes, but to an ensemble of inflating islands of finite size
inside of each bubble.
These islands are called quasi-open universes.
Within each island, the number of e-foldings of inflation
decreases as we move from the center to the edges. Also, each island
is characterized by a different number of e-foldings in its central region.
As a result, even within the same bubble, different observers will measure
a range of values of the density parameter. The picture of the large scale
structure of the universe in these models is rather simple, because
all bubbles have the same statistical 
properties.  We shall see that the quasiopen nature of inflation is of
crucial importance for the calculation of the probability distribution
for the density parameter.

In models of quasiopen inflation, such as (\ref{coupled}), $\Omega$
takes different values in different parts of the universe, while the
other constants of Nature and cosmological parameters remain fixed.
More general models can be constructed where other parameters can
change as well, and in Section VII we give an example of a
model with a variable amplitude of density fluctuations.  However, our
main focus in this paper is on the models in which only $\Omega$ is
allowed to vary.

In order to apply the principle of mediocrity to our models, we will
have to compare the number of civilizations in parts of the universe
with different values of $\Omega$.
Of course, we cannot calculate the number of civilizations.  However,
since the value of $\Omega$ does not affect the physical precesses
involved in the evolution of life, this number must be proportional to
the number of habitable stars or, as a rough approximation, to the
number of galaxies.  Hence, we shall set the probability for us to
observe a certain value of $\Omega$ to be proportional to the number
of galaxies formed in parts of the universe where $\Omega$ takes the
specified value.  

The principle of mediocrity was applied to calculate
the probability distribution for $\Omega$ in an earlier paper
\cite{VW}, which assumed the old picture of homogeneous open
universes inside bubbles.  A serious difficulty encountered in that 
calculation was 
that open universes inside the bubbles have infinite volume
and contain an infinite number of galaxies.  Thus, to find the
relative probability for different values of $\Omega$, one had to
compare infinities, which is an inherently ambiguous task.
This problem was addressed in \cite{VW} by introducing a cutoff and
counting only galaxies formed prior to the cutoff.  Although the
cutoff procedure employed in \cite{VW} has some nice properties, it is
not unique, and the resulting probability distribution is sensitive to
the choice of cutoff \cite{linde}.  This cutoff dependence, which also
appears in other models of eternal inflation \cite{linde,vireg}, has lead some
authors to doubt that a meaningful definition of probabilities in such
models is even in principle possible \cite{linde,gbl}.

However, this pessimistic conclusion may have been premature.
According to the quasiopen picture, $\Omega$ takes all its possible
values within each bubble.  Since all bubbles are statistically
equivalent, it is sufficient to consider a single bubble.  Moreover,
we can restrict ourselves to a finite (but very large) comoving volume
within that bubble, provided that its size is much greater than the
characteristic scale of variation of $\Omega$.  Thus, we no longer
need to compare infinities, and the problem becomes well defined.

The possibility of unambiguous calculation of probabilities in the
quasiopen model was our main motivation for revising the analysis of
Ref.\cite{VW}.  Also, we shall give a more careful treatment of
the astrophysical aspects of the problem which were discussed
rather sketchily in \cite{VW}.

The paper is organized as follows.
In Section II we review the main features of quasi-open inflation.
In Section III we introduce the probability distribution for $\Omega$.
A basic ingredient in this distribution
will be the anthropic factor $\nu(\Omega)$, which gives the number of 
civilizations that develop per unit thermalized volume in a region 
characterized by a certain value of $\Omega$.
In Section IV we evaluate $\nu(\Omega)$ and 
calculate the probability distribution for $\Omega$ in the model
(\ref{coupled}). 
In Section V we extend our results to more general models with arbitrary 
slow roll potentials for the field $\phi$. In Section VI we discuss
observational constraints on quasiopen models due to CMB anisotropies
and how these constraints restrict the class of models that give
a probability distribution peaked at a non-trivial value of $\Omega$.
In Section VII we comment on the ``cosmic age coincidence'', that is,
on whether it would be surprising to find ourselves living at the time when 
the curvature of the universe starts dominating. In Section VIII we 
summarize our conclusions.  Some side issues and technical details are
discussed in the appendices.

\section{Quasi-open inflation} 

In this Section we shall review the main features of quasi-open 
models which will be relevant to our discussion. To begin with,
we shall consider a model of the form (\ref{coupled}). In Section
V we shall consider more general slow-roll potentials.

As mentioned in the introduction, the interior of a bubble is isometric 
to an open Friedmann-Robertson-Walker universe, with line element 
\begin{equation}
ds^2= -dt^2 + a^2(t) [dr^2+ \sinh^2 r (d\theta^2 +\sin^2 \theta d\varphi^2)].
\label{frw}
\end{equation} 
The scale factor $a$ obeys the Friedmann equation
\begin{equation}
H^2\equiv \left( {\dot a\over a}\right)^2 = {8\pi G\over 3} \rho + 
{1 \over a^2}.
\label{friedmann}
\end{equation}
At sufficiently early times ($t\to 0$), the curvature term in the 
r.h.s. dominates over the energy density $\rho$ of the scalar fields,
and the scale factor behaves as $a \approx t$.

For the second period of slow roll inflation
inside the bubble, the energy density of
the scalar fields must be dominated by the potential term
\begin{equation}
V(\sigma,\phi) \gg \dot\sigma^2, \dot\phi^2. 
\end{equation}
Inside the bubble, the field $\sigma$ quickly settles down to 
its v.e.v. $\sigma=v$ with $V_t(v)=0$.
This happens on a time scale of order $t_0\sim M^{-1}$, where $M$ is
the typical mass scale of $V_t$. After that,
$\phi$ becomes a free field with constant mass
$$
m^2={g\over 2} v^2,
$$
and the condition for inflation becomes $\phi 
{\buildrel >\over \sim} M_p$, 
where $M_p^2=G^{-1}$.

An important feature of quasi-open models is the
existence of the so-called supercurvature modes for the slow roll field 
$\phi$. These are modes 
which are not normalizable on the infinite $t=const.$ hyperboloids
inside the bubble, but which nevertheless have to be included in
the field expansion. The reason is
that they are normalizable on the Cauchy surface where 
equal time commutation relations are imposed. Supercurvature 
modes are characterized by their eigenvalue of the Laplacian
on the 3-hyperboloid. 
For the model (\ref{coupled}), this eigenvalue is given by \cite{GGM}
\begin{equation}
\gamma = {1\over 8} H_F^2 R_0^4 m^2 \ll 1.
\label{gamma}
\end{equation}
Here,
$$
H_F^2= {8\pi G\over 3} V_0(0)
$$
is the Hubble rate during the first stage of false vacuum dominated inflation
and $R_0$ is the size of the bubble at the time of nucleation, which can
be given in terms of the model parameters \cite{parke}. Typically,
$m^2 \ll H_F^2 < R_0^{-2}$, from where the condition $\gamma \ll 1$ follows.

Around the time $t_0$ when $\sigma$ settles down to its $v.e.v.$, the 
field $\phi$ will be in a homogeneous and isotropic quantum state with 
mean squared amplitude given by \cite{GGM}
\begin{equation}
f^2 \equiv \langle\phi^2\rangle \approx \left({H_F \over 2\pi}\right)^2 
{1\over \gamma}.
\label{rms}
\end{equation}
The presence of the factor $\gamma^{-1}$ reflects the fact that $f$ is 
dominated by the contribution of supercurvature modes. Introducing
$\gamma$ from (\ref{gamma}), we find
\begin{equation}
f\approx {\sqrt{2} \over \pi} {1\over m R_0^2}.
\label{f}
\end{equation}
Up to numerical factors, this is basically the finite temperature
dispersion of a field of mass $m$ at the Rindler temperature given by
$T=(2\pi R_0)^{-1}$. 
The correlation length of $\phi$ is given by \cite{GGM}
$r \sim \gamma^{-1}$. This means that at the time
$t\sim t_0$ we can divide the space into regions of co-moving size 
$r\sim \gamma^{-1} \gg 1$ where the field is coherent.
Notice that the size of these regions is much larger than the
curvature scale $r=1$.

The parameters of the model can be chosen in such a way that $f$
is close to the Planck scale, and in that case the slow roll field
easily reaches inflating values $\phi\sim M_P$. 
For instance, if the potential $V_t$ is such that the bubble walls are
thick, then $R_0\sim M^{-1}$. Taking $M\sim 10^{16} GeV$ and 
$m \sim 10^{13} GeV$, we find $f \sim M_p$. In this case, inflating 
regions of co-moving size $r\sim \gamma^{-1}$ where the
field is large and positive will be next to inflating regions where the
field is large and negative. These two inflating regions will be
separated by regions where the field is small and the universe does not
inflate. 

The parameters can also be such that $f \ll M_p$, and in that case
most of the regions will not attain an inflating value of $\phi$.
Inflation will only happen in those regions where, as a result of
a statistical fluctuation, the field happens to be far above its
r.m.s. value. Since the volume of the hyperboloid is infinite,
there will be a small but finite density of these inflating islands
inside of each bubble. Those rare ``high peaks'' will have spherical 
symmetry. If we take the inflating patch to be centered 
at $r=0$, the radial profile of the field is given by
\begin{equation}
\phi(t_0,r) \approx \phi_0\ {\sinh[(1-\gamma)^{1/2} r]\over 
                              (1-\gamma)^{1/2} \sinh r},
\label{island}
\end{equation}
Where $\phi_0\equiv\phi(t_0,0)$ is a constant.
The probability distribution for $\phi_0$ 
is given by
\begin{equation}
P(\phi_0) \propto \exp\left[-\phi_0^2\over 2 f^2\right]
\label{pa}
\end{equation}
The variation of $\phi_0$ within the bubble results in a
position-dependent number of inflationary e-foldings, and thus in a
variable density parameter $\Omega$.  Note that all other cosmological
parameters, such as the amplitude of the density fluctuations, remain
fixed throughout the bubble (and are the same for all bubbles).
The probability (\ref{pa}) is one of the basic ingredients from which the 
most probable value of $\Omega$ is calculated.

It should be mentioned that the size of an inflating region
can be much larger than the size of the actual ``populated'' region
within it, $r_p$, where matter will cluster efficiently into galaxies.
The size of the populated region is calculated in Appendix A. 
This size should be larger than the present horizon, since otherwise
we would observe large anisotropies in the galaxy distribution.  For
$\Omega$ not too close to 1, the horizon distance is comparable to the
curvature scale, $r=1$, and we have to require that $r_p > 1$.  The
corresponding constraint on $\gamma$ is obtained in Appendix A.

Equation (\ref{pa}) can be understood from a different perspective,
by using the Euclidean approach to the calculation of the 
nucleation rate. The strategy is to study how this rate
is affected by the local value of $\phi$ at the place where the bubble
nucleates. This is simple because we only need this 
Euclidean action to quadratic order. Taking $\phi=0$, we
denote by $\sigma_0(\tau)$ the $O(4)$ symmetric instanton \cite{coleman}
responsible for vacuum decay. Here, $\tau=it$  is the Euclidean
time, the ``radial'' coordinate on which the instanton depends. 
Expanding the Euclidean action $S_E$ to second order in
perturbations of $\sigma$ and of $\phi$, the perturbations
in $\sigma$ and $\phi$ will decouple to quadratic order.
Taking $\phi=\phi_0=const.$ the change in the
Euclidean action will simply be
\begin{equation}
\Delta S_E = \int (g/2) \sigma_0^2(\tau) \phi_0^2 d^4 x         
\label{shortcut}
\end{equation}
We can approximate the integral by taking $g \sigma^2_0= m^2$
inside the volume of the bubble and $\sigma_0=0$ outside. Then we have
$$
\Delta S_E = {\pi^2 \over 2} m^2 R_0^4  {\phi_0^2\over 2}.
$$
From the formula $P\sim \exp(-S_E)$, we essentially recover Equation
(\ref{pa}). Even though we have used the thin wall approximation, 
we should stress
that the coincidence of this ``adiabatic'' result (where the 
field $\phi$ is taken as constant) with the field theoretic
one (where $\phi$ is quantized in the bubble background and its
r.m.s. is evaluated right after bubble nucleation)
is also valid for thick walls \cite{GGM}.

As emphasized in \cite{GGM}, the adiabatic approach to the calculation
of the distribution of $\phi_0$ should be interpreted with caution. It
does not mean that the surface $t=t_0$ inside the bubble will have 
a constant value of the field $\phi$. It only gives the probability
that a bubble will nucleate with the value of $\phi=\phi_0$ near 
$r=0$. We know, however, that the quantum state of
a nucleating bubble is homogeneous, and therefore in
the ensemble of bubbles there is nothing special 
about the point $r=0$. Therefore, 
this also gives the probability distribution for $\phi$ around any point  
inside the bubble.

\section{Probability distribution for $\Omega$}

In this Section we shall follow some of the steps used in
Ref. \cite{VW} for the calculation of the probability
distribution for $\Omega$, although the present case will actually be simpler. 
In the case of Ref. \cite{VW},
one had to deal with an infinite number of bubbles, each one
containing an infinite open universe with a different density
parameter. Since the probability for a given set of parameters is
roughly proportional to the total volume that ends up having those
values of the parameters, one had to face the difficulty
of comparing infinite volumes in an eternal inflationary universe
\cite{vireg}.

In our case, all bubbles are statistically equivalent. All
of them are described by a homogeneous and isotropic quantum state,
with $\langle\phi^2\rangle$ given by (\ref{rms}). 
Hence, in order to find
the probability distribution for $\Omega$ it is sufficient to
look at the interior of a single bubble. Also, since the quantum 
state is homogeneous, we only need to consider the evolution of 
a patch of finite co-moving size around an arbitrary point on the  
$t\sim M^{-1}$ hyperboloid. The patch should be sufficiently large
that it contains regions with all possible values of $\phi$, distributed
according to (\ref{pa}).
Since inflation inside the bubble is
not eternal, the number of civilizations resulting from this
co-moving patch is finite and there is no need for regularization.

As mentioned in the previous section, at early times the scale factor
behaves as $a \approx t$. By the time $t\sim H^{-1}(\phi_0)$,
where 
$$ H(\phi_0)= (4 \pi G/3)^{1/2} m \phi_0, $$
the energy density in the scalar field $\phi$ starts dominating over 
the curvature term. If the condition for slow roll inflation
\begin{equation}
\phi_0> \phi_{th} \equiv {M_p\over \sqrt{4\pi}}
\label{phistar}
\end{equation}
is satisfied, then using equation (\ref{friedmann}) the scale factor will 
subsequently evolve as
$$
a(t) \approx  H^{-1}_0  e^{N(t)},
$$ 
where $H_0 \equiv H(\phi_0)$ and
$$
N(t) = \int_{\phi_0}^{\phi(t)} {H(\phi)\over \dot\phi} d\phi.
$$
Using the slow-roll equation of motion for $\phi$
$$
\dot \phi ={-m^2\over 3H} \phi,
$$
we have $N(t) \approx 2\pi G\ [\phi_0^2-\phi^2(t)]$.
Since $\phi_0$ is actually a slowly varying function of
position,  the scale factor is a local one, and should be understood
as $a(t,x^i)$. Notice that the co-moving scale over which $a$ changes is 
comparable to $\gamma^{-1}$ and hence it is much larger than the
curvature scale, so
it is meaningful to use the Friedmann equation (\ref{friedmann}).

The number of e-foldings of inflation depends on the local value
of $\phi_0$
\begin{equation}
a_{th}(\phi_0)\equiv H^{-1}_0 e^{N_{th}(\phi_0)}\approx  
                     H^{-1}_0 e^{2\pi G (\phi_0^2-\phi_{th}^2)},
\label{efoldings}
\end{equation}
where $\phi_{th}$ is defined in (\ref{phistar}).
It will be convenient, as a first step, to find the
probability distribution for a random ``civilization'' to live in a region
which had a value of the slow roll field equal to $\phi_0$ at the beginning
of inflation. This is given by
\begin{equation}
d {\cal P}(\phi_0) = P(\phi_0) a^3_{th}(\phi_0) \nu(\phi_0)
             d \phi_0.
\label{calp}
\end{equation}
Here 
\begin{equation}
P(\phi_0) \propto e^{-\phi_0^2 \over 2f^2}
\label{gauss}
\end{equation}
is the probability that a 
given point on the $t\sim M^{-1}$ hyperboloid will have the value $\phi_0$
right after nucleation. 
Because the number of civilizations is proportional to the volume, we 
have inserted the total expansion factor during inflation $a^3_{th}(\phi_0)$.
Finally, $\nu(\phi_0)$ is the ``human factor'', which
represents the number density of civilizations that will develop
per unit thermalized volume as a function of $\phi_0$. 

Factoring out the dependence on $\nu$,
\begin{equation}
d {\cal P} =  \nu(\phi_0) d\tilde{\cal P},
\label{tildep}
\end{equation}
the leading exponential behaviour of $d\tilde{\cal P}$ is  
$$
\exp \left(6\pi G -{1\over 2 f^2}\right) \phi_0^2,
$$
where we have used (\ref{efoldings}) and (\ref{calp}).
The behaviour of $\tilde{\cal P}$ depends on whether $f$ is large
or small compared with $M_p$. Defining
\begin{equation}
\mu\equiv {1\over 24 \pi G f^2},
\label{mu}
\end{equation}
it is clear that for $\mu<1/2$ large values of $\phi_0$
are favoured due to the gain in volume factor, and we may expect the
universe to be very flat. For $\mu>1/2$, the volume factor alone
is not sufficient to compensate for the exponential suppression of
high peaks. We shall see that the human factor may play an important role
in this case .

It is convenient
to express the above distribution in terms of the density parameter.
Following \cite{BGT,VW}, we have
\begin{equation}
[H(\phi_{th}) a_{th}(\phi_0)]^2= 1+ B {\Omega \over 1-\Omega} \approx 
B {\Omega \over 1-\Omega}.
\label{omegaphi}
\end{equation}
Here 
$$
B \approx {T_{th}^2 \over T_{eq} T_{CMB}},
$$
$T_{eq}$ is the temperature at equal matter and radiation density,
$T_{th}$ is the thermalization temperature and $T_{CMB}$ is the 
temperature of the cosmic microwave background, measured at the 
same time as $\Omega$.
Typically,
$B$ is exponentially large, with $(\ln B) \sim 10^2$.
>From (\ref{calp}) and (\ref{omegaphi}) we find
\begin{equation}
d \tilde{\cal P} (\Omega) = P(\phi_0) 
                     a^3_{th}\ 
\left({d \ln a_{th} \over d \phi_0}\right)^{-1}\  
                     {d \Omega \over 2 \Omega (1- \Omega)}.
\end{equation}                             
Using (\ref{efoldings}), and disregarding the
logarithmic dependence on $\Omega$, we find
\begin{equation}
d \tilde{\cal P} (\Omega) 
\propto \Omega^{1/2-3\mu} (1-\Omega)^{3\mu - 5/2} d\Omega,
\label{19}
\end{equation}
where $\mu$ is given by Eq. (\ref{mu}). 
For $\mu>5/6$ the probability 
distribution
is peaked at $\Omega=0$, for $\mu<1/6$ it is peaked at $\Omega=1$,
and for the intermediate range $1/6 <\mu< 5/6$ it has 
two peaks, one at $\Omega=0$ and one at $\Omega=1$. However, it is
easily seen that for $\mu>1/2$ the highest peak will be at $\Omega=0$
whereas for $\mu<1/2$ it will be at $\Omega=1$.\footnote{Strictly speaking, 
the peak would not be exactly at $\Omega=1$ because the Gaussian
distribution (\ref{gauss}) is only an approximation which 
ignores the backreaction of the slow roll field on the bubble background. 
We shall return to this issue in Section V.}   Note that all
dependence on the particle physics model in Eq.(\ref{19}) has been
compressed into a single parameter $\mu$.

Eq. (\ref{19}) is the same expression that was found in Ref. \cite{VW}
by considering an ensemble of bubbles with different values of $\Omega$
and using the prescription introduced in \cite{vireg} for the regularization
of infinite volumes. We regard the agreement between the two approaches as
a validation of this regularization prescription (in models where
regularization is needed). Alternative 
regularizations proposed in \cite{linde} give different results 
and are therefore disfavoured.

Let us now include the human factor $\nu(\Omega)$. As mentioned above,
this will play a role for $\mu>1/2$, when the expansion
alone is not sufficient to compensate for the exponential
suppression in $\phi_0$ due to tunneling. 
Since the probability distribution 
$\tilde {\cal P}$ tends to peak near the extremes, it
is convenient to work with a logarithmic variable which gives equal measure
to each decade in the vicinity of $\Omega=0$ or $\Omega=1$. One such 
variable is 
$\ln x$, where
\begin{equation}
x\equiv {1-\Omega \over \Omega}.
\label{lx}
\end{equation}
Hence, we shall be interested in the probability density
\begin{equation}
W(\Omega)={d{\cal P} \over d\ln x}\propto \nu(x)\ x^{3(\mu-1/2)}.
\label{peak}
\end{equation}
The peak of this distribution will give the most probable value
of $\Omega$. 

It should be noted that, since the density 
parameter changes with time, both $\nu(\Omega)$ and $W(\Omega)$
are in principle time dependent. However, this time dependence is
somewhat trivial, entering (\ref{peak}) through the parameter $T_{CMB}$, 
the temperature at which the density parameter is equal to $\Omega$.
What we are actually interested in is the probability distribution
for different types of thermalized regions, which is intrinsically time 
independent. We could, for instance, 
set $T_{CMB}$ equal to the temperature at recombination, and then the 
probability distribution would be expressed in terms of $\Omega_{rec}$,
which completely characterizes the history of a given thermalized region.
Noting that Friedmann's equation can be rewritten 
as $x^{-1}=(8\pi G/3) \rho a^2$, in the matter era we have 
\begin{equation}
x \propto a(t) \propto T_{CMB}^{-1}, \label{xprop}
\end{equation}
where $a$ indicates the scale factor. Hence, in practice, we can use 
a ``gauge invariant'' approach: we shall write
$W(\Omega)$ as a function of the product $x\ T_{CMB}$, which is time 
independent in the matter era.

\section{The anthropic factor $\nu(\Omega)$}

In previous work \cite{ef96,V96,W,VW,MSW}, $\nu(\Omega)$
was taken to be proportional to the fraction of clustered matter $f_c$
on a relevant mass scale $M_g$. This scale can be chosen as the typical 
mass of an $L_*$ galaxy, $M_g \sim 10^{12}M_{\odot}$ \cite{ef96,MSW},
given that most of the observed luminous 
matter is in this form.
Also,
galaxies much smaller than $10^{12}M_{\odot}$ may not be suitable for 
life, because their gravitational potential would not be able to
hold the heavy elements produced in supernovae explosions.
Matter will only cluster when the density contrast $\delta(M_g)$
extrapolated from linear perturbation theory  
exceeds a certain threshold $\delta_c$. 
Hence $f_c$ can be estimated as
\cite{PS74,lid95}
\begin{equation}
f_c(M_g, t) = {\rm erfc}\left({\delta_c \over \sqrt{2}\sigma}\right).
\label{fraction}
\end{equation}
Here ${\rm erfc}$ is the complementary 
error function and $\sigma(M_g, t)$ is the dispersion
in the density contrast, also evolved according to linear theory
\footnote{This expression for the
growth of perturbations is different from the one used in \cite{VW}. There,
the growth factor from the time of equilibrium of matter and radiation was
considered, and a spurious factor of $\Omega$ was included, which was 
actually due to the uncertainty in the value of the redshift at the time 
of equilibrium. This factor should actually not be present in the 
probability distribution for $\Omega$, since the time of equilibrium 
is the same in all thermalized regions.}
\cite{peebles}
\begin{equation}
\sigma(M_g,t)= {5\sigma_{rec}(M_g) \over 2 x_{rec}} f(x),
\label{growth}
\end{equation} 
where $x$ is given by (\ref{lx}) and
\begin{equation}
f(x) = 1+{3\over x}+{3(1+x)^{1/2} \over x^{3/2}} \ln[(1+x)^{1/2} -x^{1/2}].
\label{function}
\end{equation}
The subindex $rec$ denotes quantities evaluated at the time of 
recombination. In an open universe, perturbations stop
growing after the universe becomes curvature dominated. Since 
we are interested in the total fraction of clustered matter in the
entire history of a given region, we should use in (\ref{fraction})
the asymptotic value of $\sigma$ at large times ($x\to \infty$), which 
approaches a constant.

In a flat universe, the critical density contrast takes the value
$\delta_c \approx 1.7$. However, it is known that
$\delta_c$ should be slightly $\Omega$-dependent \cite{deltac}. The
variation is rather small, and $\delta_c$ changes by no more than 5\% as
$\Omega$ varies from 0.1 to 1. Here we adopt the value of $\delta_c$ 
estimated in the spherical collapse model as \cite{laco93,foot1}
\begin{equation}
\delta_c(x) = {3\over 2} f(x) g(x),
\label{deltac}
\end{equation}
where
\begin{equation}
g(x) \equiv 1+
 \left({\pi \over x^{1/2} (1+x)^{1/2} - \sinh^{-1}x^{1/2}}\right)^{2/3}.
\label{g}
\end{equation} 
For $x \to 0$, we have $\delta_c=(3/5)(3\pi/2)^{2/3} \approx 1.69$, as in 
the case of a flat universe, and for $x\to \infty$ we have $\delta_c = 3/2.$

Substituting (\ref{growth}) and (\ref{deltac}) in Eq. 
(\ref{fraction}) and taking the limit $x\to \infty$ we obtain
\begin{equation}
\nu = {\rm erfc}\left({3 x_{rec}\over 5 \sqrt{2} \sigma_{rec}}\right) 
\equiv {\rm erfc}(y).
\label{erfc1}
\end{equation}
The distribution (\ref{erfc1}) is given as a function of the density
parameter at the time of recombination. As mentioned at the end of the
last section, in order to compare predictions with observations, it is
convenient to express the distribution as a function
of $x$ at any temperature $T_{CMB}$. Using (\ref{xprop}) we have
\begin{equation}
y=\kappa x \equiv {3 \over 5 \sqrt{2} \sigma_{rec}} {T_{CMB} \over T_{rec}} x.
\label{kappa}
\end{equation}
In order to evaluate the coefficient $\kappa$, we need to  
know $\sigma_{rec}$.
It is clear that $\sigma_{rec}$ has nearly the same value in all regions where 
curvature dominates only well after the time of recombination. In principle 
this value is given in terms of the parameters of our theory of initial 
conditions.  

\begin{figure}[t]
\centering
\hspace*{-4mm}
\leavevmode\epsfysize=10 cm \epsfbox{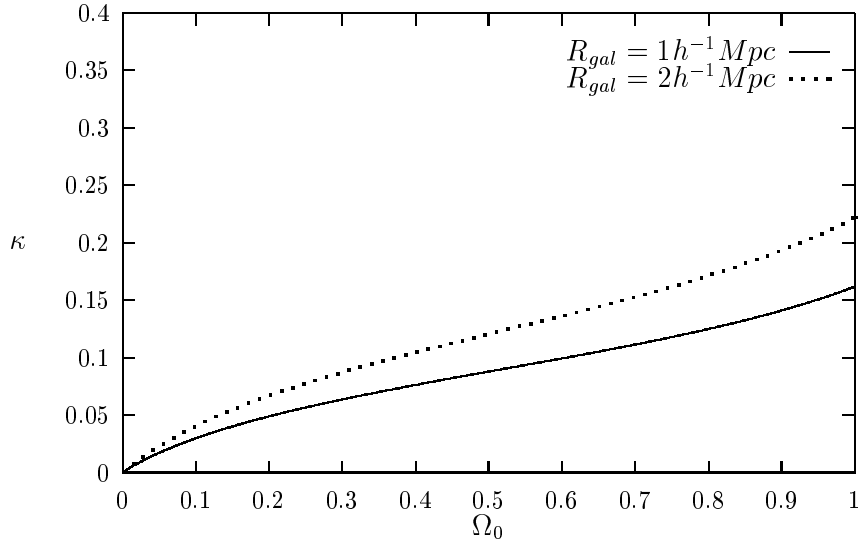}\\[3mm]
\caption[fig1]{\label{fig1} The coefficient $\kappa$
which relates the variable $y$ to the density parameter
$y= \kappa x = \kappa (1-\Omega)/\Omega$, depends on 
$\sigma_{rec}$, the
value of the density contrast at the time of recombination.
Our ability to infer $\sigma_{rec}$ from CMB observations is
limited by the fact that $\Omega_0$ in our observable ``subuniverse''
is not known very precisely. In the figure we plot the inferred
value of $\kappa$ for various assumed values of $\Omega_0$.
The value of $\sigma_{rec}$ depends moreover on the scale $R_{gal}$
corresponding to objects of galactic mass. The curve is plotted for 
two different values  of this scale (see Appendix B). The parameter
$\kappa$ depends on the temperature at which we observe $\Omega$. 
Here we have taken $T_{CMB} = 2.7 K$ .}
\end{figure}

In practice, we can adjust the parameters of the theory to fit 
CMB observations. 
Our ability to infer $\sigma_{rec}$ from CMB observations is, however, 
limited by the fact that it depends on the 
values of $\Omega_0$ and $h$ in our visible universe, which are not very well
determined.  As noted in \cite{MSW},
this limitation also arises in attempts to find the probability
distribution for the cosmological constant.
Therefore, until determinations of $\sigma_{rec}$ become more precise,
the best one can do is to assume certain values of 
$\Omega_0$ and $h$ and check whether the assumed values fall
within the range favoured by the resulting probability
distribution for $\Omega$. The value of $\kappa$ for $T_{CMB}=2.7 K$
is estimated in Appendix B and plotted
in Fig. 1 as a function of $\Omega_0$. For each value of $\Omega_0$,
$h$ has been chosen so that the ``shape parameter'' 
$\Gamma \approx \Omega_0 h \approx .25$ (see Appendix B). Also, there
is some uncertainty in the relevant co-moving scale $R_{gal}$  
corresponding to $M_{gal}$ \cite{MSW}. 
In the figure we consider two possibilities,
$R_{gal}= 1\ h^{-1} Mpc$ and $R_{gal}=2\ h^{-1} Mpc$.
For $\Omega_0$ in the range $0.1<\Omega_0<0.7$ we find that $\kappa\sim 0.1$.

\begin{figure}[t]
\centering
\hspace*{-4mm}
\leavevmode\epsfysize=10 cm \epsfbox{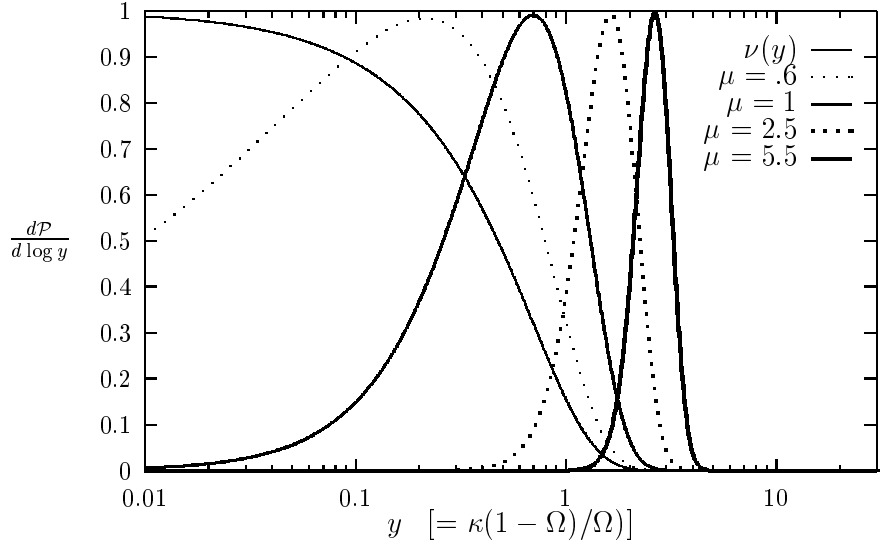}\\[3mm]
\caption[fig2]{\label{fig2} The probability distribution 
(\ref{distributiony})
as a function of $y$, for various values of $\mu$.
Also represented is the fraction of clustered matter $\nu(y)$
as a function of $y$.}
\end{figure}

The fraction of clustered matter $\nu$ is shown by a solid curve in 
Fig. 2 as a function of $y$. In our universe, the density of 
matter presently clustered in giant galaxies satisfies  $\Omega_{gal} >0.05$
\cite{peebles93}, which implies $\nu(y)>0.05\Omega^{-1}$. 
The asymptotic value $\nu(y)$ should be even larger.  Solving for $y$,
we obtain the 
observational constraint
\begin{equation}
y<0.9\ . \label{constrainty}
\end{equation}

The distribution
\begin{equation}
W(\Omega)={d {\cal P} \over d\ln y} \propto {\rm erfc}(y) y^{3(\mu-1/2)}
\label{distributiony}
\end{equation}
gives the probability that a randomly selected civilization is located in a
region which had a specified value of $\Omega$ at a given temperature 
$T_{CMB}$. It is represented in Fig. 2, as a function of $\ln y$,
for different values of the parameter $\mu$. 
The peak value $y_{peak}$, found from $dW/dy=0$ is plotted in Fig. 3 as a 
function of $\mu$ (curve $a$).

\begin{figure}[t]
\centering
\hspace*{-4mm}
\leavevmode\epsfysize=10 cm \epsfbox{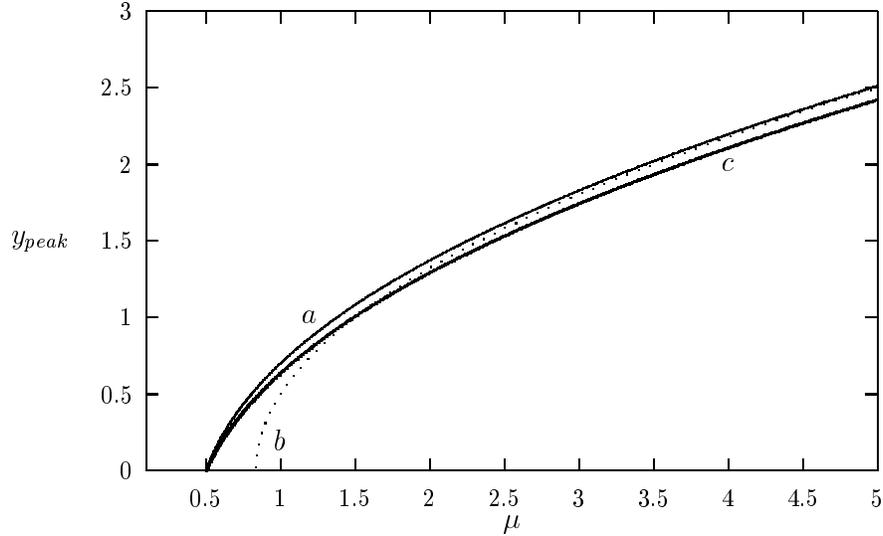}\\[3mm]
\caption[fig3]{\label{fig3} Peak of the probability distribution 
(\ref{distributiony}) (curve $a$). The approximate value of 
$y_{peak}$ given by (\ref{ypeak}) is represented by the curve $b$.
Curve $c$ represents the possible effect of helium line cooling failure, 
as discussed in Appendix C.}
\end{figure}

For $y \gtrsim 1$, the error function can be approximated by
\begin{equation}
\label{approxer}
{\rm erfc}(y)\approx {1\over \sqrt{\pi} y} e^{-y^2},
\end{equation}
and the peak value can be expressed analytically,
\begin{equation}
y_{peak}^2 \approx {3\over 2}\mu-{5\over 4}. 
\label{ypeak}
\end{equation}
This curve is also shown in Fig. 3 (curve $b$). 
Eq. (\ref{ypeak}) can be rewritten as
\begin{equation}
\left({1-\Omega\over\Omega}\right)_{peak} = 
\kappa^{-1}\left({3\over 2} \mu-{5\over 4}\right)^{1/2},
\label{omegapeak}
\end{equation}
which gives the peak value for the density parameter at the temperature
$T_{CMB}$. To estimate the width of the distribution (\ref{distributiony})
we expand $\ln W$ to quadratic order in
$\Delta \ln y$ around $y_{peak}$,
$$
W \approx W_{peak} \exp\left[{- (3\mu-{5/2}) (\Delta \ln y)^2}\right].
$$ 
Hence, the root mean squared dispersion in $\Omega$ around its peak
value will be given by (for $\mu\gtrsim 3/2$)
\begin{equation}
\Delta \ln\left({1-\Omega \over \Omega}\right) \sim (6\mu - 5)^{-1/2},
\label{deltaomega}
\end{equation}
while the dispersion in $y$ is independent of $\mu$, $\Delta y \sim 1/2$.

From Fig. 2, we see that
as $\mu$ is increased, the probability distribution is sharper
and displaced towards larger values of $y$, in agreement with
(\ref{ypeak}) and (\ref{deltaomega}). 
For $\mu=1$, the
distribution has a substantial overlap with the region where
(\ref{constrainty}) is satisfied and the fraction of clustered matter
is compatible with observations. For $\mu=5/2$ this
overlap is smaller, but still non-negligible. However, for $\mu=11/2$, 
the probability density at the point $y \approx 0.9$ is more than two orders 
of magnitude smaller than at its peak value. Particle physics
models which give such high values of 
$\mu$ are therefore disfavored by observations.
As we discussed in Section III, the probability distribution is 
peaked at $\Omega=1$ for $\mu<0.5$. Hence the range of $\mu$ that is
of interest to us in these paper is 
\begin{equation}
0.5 < \mu \lesssim 3 \label{rangemu}
\end{equation}
This
corresponds to 
\begin{equation}
0<y_{peak} \lesssim 2.
\label{o}
\end{equation}
It should be noted that the peak value for the fraction of
clustered matter $\nu(y_{peak})$, depends only on $\mu$,
and not on the primordial spectrum of density fluctuations. 

So far we assumed that all galactic-size objects collapsing at
any time will form luminous galaxies.  However, this is not
necessarily the case.  Galaxies forming at later times have lower
density and shallower potential wells.  They are vulnerable to losing
all their gas due to supernova explosions \cite{TR}.  Moreover, a
collapsing cloud will fragment into stars only if the 
cooling timescale of the cloud $\tau_{cool}$ is smaller than the
collapse timescale $\tau_{grav}$,
otherwise the cloud would stabilize into a pressure 
supported configuration \cite{reos,TR}.  The cooling rate of such 
pressure supported clouds is exceedingly low, and it is possible that
star formation in the relevant mass range will be suppressed in these
clouds even when they eventually cool.  
Hence, it is conceivable that galaxies that fail to cool during
the initial collapse give a negligible contribution to $\nu(\Omega)$
\cite{TR}.  The possible effect of cooling failure and related phenomena
on the probability distribution for $\Omega$ is discussed in Appendix
C, where we show that the effect is to  shift the peak of the
distribution towards larger values of $\Omega$.These effects may be 
significant, but not dramatic, and (\ref{omegapeak}) remains valid by 
order of magnitude. As an illustration, curve 
$c$ of Fig. 3 shows the peak of the modified distribution when
matter which clusters after the time when helium line cooling becomes 
inefficient is excluded from the anthropic factor $\nu(\Omega)$. 

\section{More general models}

In this section, we shall generalize our results to models where 
the slow roll potential is not
necessarily quadratic in $\phi$. In this case, the factor
$P(\phi_0)$ in Eq. (\ref{calp}) can be estimated in the adiabatic 
approximation, where the field $\phi$ is treated as a constant
during tunneling, as described at the end of Section II. In this
approximation we have
$$
P(\phi_0) \propto e^{-S_E(\phi_0)}
$$
where $S_E(\phi_0)$ is the action of the instanton for bubble nucleation,
with the slow roll field frozen to the value $\phi_0$.

>From (\ref{calp}), we have
\begin{equation}
W={d{\cal P}\over d\ln y} \propto {\rm erfc}(y) y^{-3/2} P(\phi_0) J^{-1},
\label{pgeneral}
\end{equation}
where, as before
$$
y= {\delta_c \over \sqrt{2} \sigma_{rec}} 
{2\over 5}{T_{CMB} \over T_{rec}} {1-\Omega \over \Omega},
$$
and we have used (\ref{omegaphi}) to express the scale factor as 
a function of $y$. The Jacobian $J$ is given by
\begin{equation}
J\equiv \left| {d\ln y\over d\phi_0} \right|
=2 {d\ln a_{th} \over d\phi_0}=-{V'\over V} + 16\pi G {V\over V'},
\label{jacobian}
\end{equation}
where we have used $a_{th}=H(\phi_0) e^{N}$, and the relation between
the hubble rate $H(\phi_0)$ and the slow roll potential 
$V$ in true vacuum
$$
H_T^2\equiv H^2(\phi_0) \approx {8\pi G\over 3} V(\sigma_T, \phi_0).
$$
Here $\sigma_T$ is the value of the tunneling field in true vacuum.
We have also used the slow roll expression for the number of e-foldings
$$
N(\phi_0)= 8\pi G \int^{\phi_0} {V\over V'} d\phi.
$$
Here, as in (\ref{jacobian}), $V'$ stands for the derivative of $V$
with respect to the slow roll field.
Introducing the slow roll parameter
\begin{equation}
\epsilon\equiv {-\dot H_T\over H_T^2}
\approx {1\over 16\pi G}\left({V'\over V}\right)^2 \ll 1,
\label{epsilon}
\end{equation}
we have 
$$
J \approx \left({16\pi G \over \epsilon}\right)^{1/2}.
$$
In many models, the parameter $\epsilon$ hardly changes in the
relevant range of $\phi_0$, and hence we shall treat it as a 
small constant parameter.

Extremizing (\ref{pgeneral}), we find that the peak value of $\phi_0$
is given by the condition
\begin{equation}
\mu(\phi_0)|_{peak}= 
{1\over 2}-{1\over 3}\left.{d\ln{\rm erfc}(y) \over d\ln y}\right|_{peak}.
\label{conditionmu}
\end{equation}
where
\begin{equation}
\mu(\phi_0)\equiv {1\over 3} {d S_E(\phi_0) \over d\phi} J^{-1}
\label{defmu}
\end{equation}
Eq. (\ref{conditionmu}) is the same condition we found in Section IV, 
and which is plotted
in Fig. 3, except that now $\mu$ is a function of $\phi_0$, and hence of
$y$.

Before we proceed, let us go back to the case discussed in Section III of a 
free slow roll field . Strictly speaking, the expression 
(\ref{gauss}) for $P(\phi_0)$ is just an approximation which is valid
only for sufficiently low $\phi_0$, when the backreaction
of $\phi$ on the bubble background can be neglected. Now we can take 
this effect into account. For definiteness, 
let us consider the case where 
$m \sim 10^{13} GeV$, and where the tunneling potential 
$V_t(\sigma)$ is such that false and 
true vacuum are strongly non-degenerate when $\phi=0$. In this case, 
the radius of the bubble is $R_0\sim M^{-1}$ (thick wall bubble),
where $M\sim 10^{16} GeV$ is a typical mass scale in the tunneling potential.
Let us denote by $\phi_{deg}$ the value  
for which the energy density corresponding to the slow roll potential 
is equal to the false vacuum energy in the unbroken phase 
$V_F\equiv V_t(\sigma=0)$,
$$
{1\over 2} m^2 \phi_{deg}^2 = V_F \sim {M^4\over \lambda},
$$
where $\lambda$ is a self-coupling of the tunneling potential.
For $\phi_0 \ll \phi_{deg}$, the value of $\mu$ is almost independent
of $\phi_0$ (this is the situation considered in Section III)
$$
\mu(\phi_0\ll \phi_{deg})\approx 
\mu_0\equiv {\pi \over 48 G}{m^2\over M^4}.
$$
The masses $M$ and $m$ can be easily adjusted so that $\mu_0\ll 1/2$.
However, for $\phi_0 \sim \phi_{deg}$, the Euclidean action $S_E(\phi_0)$ 
increases very steeply with $\phi_0$, and so does $\mu$
\footnote{Indeed, as $\phi_0$ approaches $\phi_{deg}$, the thin wall
approximation starts to apply. Then, from Eqs (\ref{rnot}) and (\ref{snot})
below, we find that the action blows up as we approach degeneracy}. 
In this case, the condition (\ref{conditionmu}) will be satisfied 
for $\phi_0 \sim \phi_{deg} \sim M_p (\lambda\mu_0)^{-1/2}$, where 
$M_p=G^{-1/2}$ 
is the Planck mass. The corresponding number of e-foldings of inflation 
is given by 
$N(\phi_0)\approx 2\pi G \phi_{deg}^2 \approx 
(\pi^2/ 12)(\lambda\mu_0)^{-1}$. 

Therefore, for $\mu_0\ll 1/2$, and with a suitable choice of $\lambda$,
the peak in the distribution may be adjusted to correspond to $N\approx
(1/2) \ln B \approx 60$, where $B$ is defined in (\ref{omegaphi}).
This is compatible with an open 
universe. 
However, this case is somewhat trivial, in the sense that
the universe can be open only if the maximum allowed value of the
slow roll field after tunneling, $\phi_0=\phi_{deg}$, does not drive 
a long enough period of inflation to make it flat.

Turning to the general case,
a more interesting situation arises when $\mu(\phi_0) >1/2$ throughout
the range of $\phi_0$ (see Section III). In this case the product of 
tunneling and volume factors would peak at 
$\phi_0=0$, where the resulting universe would be almost empty, and
the anthropic factor $\nu(\Omega)$ is crucial in determining 
the probability distribution for $\Omega$. For large $\mu$, and 
using the approximate expression (\ref{approxer}) for the error function
in (\ref{conditionmu}) we have
$$
y_{peak}^2\approx {3\over 2}\mu(\phi_0)-{5\over 4},
$$
which is formally the same expression as (\ref{ypeak}).

In the thin wall approximation, we can estimate $\mu$ in terms of
$V$ and the bubble radius. For simplicity, we shall also neglect 
gravitational backreaction. Denoting by $S_1$ the tension of the bubble 
wall, the radius of the bubble at the time of nucleation is given by 
\cite{coleman}
\begin{equation}
R_0={3 S_1 \over \Delta V(\phi_0)},
\label{rnot}
\end{equation}
where $\Delta V\equiv  V_F-V(\sigma_T,\phi_0)$. Here
$V_F$ is the potential in false vacuum.
For our approximation to be valid, $R_0$ should be larger than
the thickness of the bubble wall and smaller than the Hubble 
radius in false vacuum.
Under these assumptions, the Euclidean action is given by \cite{coleman}
\begin{equation}
S_E\approx {\pi^2\over 2} S_1 R_0^3.
\label{snot}
\end{equation}
The derivative of $S_E$ can be expressed in
terms of the slow roll parameter
$$
{dS_E \over d\phi} = 3 S_E {V' \over \Delta V(\phi_0)}=
{\pi^2 \over 2} R_0^4 V' = 
{\pi^2 \over 2} R_0^4 V (16 \pi G\epsilon)^{1/2},
$$ 
and finally, from (\ref{conditionmu}), we have
\begin{equation}
\mu= {\pi^2\over 6} R_0^4 V \epsilon.
\label{mugeneral}
\end{equation}
Taking one more derivative of $\ln W$ with respect to $\ln y$, we find
$$
{d^2 \ln W\over d(\ln y)^2} \approx -4 y^2 
-{\pi^2 \over 2} R_0^4 V \epsilon^2 \left(1+4{V\over V_F-V}\right).
$$
Near $y=y_{peak}$ we have, setting the first derivative of $W$ to zero and
using $\epsilon \ll 1$,
$$
\left.{d^2 \ln W\over d(\ln y)^2}\right|_{peak} 
\lesssim -4 y_{peak}^2.
$$
From this we can estimate the dispersion in the distribution of 
$\Omega$, which is again approximately given by (\ref{deltaomega}).

\section{Constraints from CMB anisotropies}


As we have shown, given a particle physics model which leads to
quasi-open inflation, we can predict the  probability distribution 
$P(\Omega)$. Of course, the model also makes predictions for
the CMB anisotropies. Comparison of all predictions
with observations can be used to constrain the parameters 
of the particle physics model.

In an open (or quasi-open) universe, 
CMB anisotropies which are generated during inflation
come in three different types. The first
type corresponds to scalar fluctuations generated during
slow roll inside the bubble, and it affects wavelengths smaller
than the curvature scale. These are called subcurvature modes. 
The corresponding spectrum of temperature fluctuations, characterized
by the multipole coefficients $l(l+1) C_l$ as a function of $l$,
is nearly flat for $l\lesssim 100$.
This type of fluctuations is usually believed to give the dominant 
contribution to the observed plateau in the CMB spectrum.

The second type of anisotropies corresponds to excitations of $\phi$ 
generated outside the bubble or during the process of tunneling and 
expansion of the bubble
into the false vacuum. These are accounted for by the supercurvature
modes discussed in section II (see also \cite{GGM}). 
For the models we have considered, the 
amplitude of temperature anisotropies caused by supercurvature modes 
is a factor 
of order $H_F/10 H(\phi_0)$ relative to the subcurvature ones \cite{tama}. 
However,
supercurvature modes affect only the very few first multipoles, and hence 
they cannot explain the observed flat spectrum. For that reason,
the constraint $H_F \lesssim 10 H(\phi_0)$ is usually imposed.  

Finally, there are CMB anisotropies caused by gravity waves,
which can in turn be decomposed into the ones generated during 
slow-roll and the ones caused by fluctuations of the bubble wall itself
\cite{tama,GMST}. Wall fluctuations give the dominant contribution 
for the few first multipoles, but their contribution decays rapidly
with $l$. The waves generated during slow roll give an approximately
flat spectrum, whose amplitude is much smaller than that of
scalar modes.

The multipole coefficients $C_l$ for the temperature 
anisotropies due to subcurvature modes are given by 
\cite{tama,juanopen}
\begin{equation}
D_l^S \equiv {l(l+1) C_l^S \over 2\pi} =
{4\pi G\over 25}\left({H_T\over 2\pi}\right)^2 {1\over \epsilon} 
b_l(\Omega). \quad (l\lesssim 100)\label{deltas}
\end{equation}
Here, we have used the notation $H_T\equiv H(\phi_0)$
and the slow roll parameter $\epsilon$ given in (\ref{epsilon}).
The coefficient $b_l$ is a slowly varying function of $\Omega$ which 
can be bounded as 
$1\lesssim b_l \lesssim 6$ in the range $.1<\Omega<1$.

Supercurvature modes induce 
temperature anisotropies which for the lowest multipoles can be estimated 
as \cite{tama,juanopen}
\begin{equation}
D_l^{SC} \equiv {l(l+1) C_l^{SC} \over 2\pi} \sim 
d_l \left({H_F\over H_T}\right)^2 D_l^S \label{deltasc}
\end{equation}
where $d_l(\Omega) \sim 10^{-2}$.

Let us now consider the contribution to CMB anisotropies from 
tensor modes. As mentioned above, for the lowest multipoles
this is dominated by the domain 
wall fluctuations\cite{ga,tama,juanopen}. For simplicity, we shall consider 
the case of a weakly gravitating domain wall, satisfying $G S_1 R_0 \ll 1$,
where $S_1$ is the wall tension. Also, we shall restrict attention to 
the thin wall case. Then, the anisotropies caused by the wall 
fluctuations are given by \cite{juanopen,ga,tama}
\begin{equation}
D_l^W\approx {2 H_T^2 \over \pi S_1 R_0} c_l(\Omega).
\end{equation}
For the first few multipoles, and $\Omega$ in the 
range $.1$ to $.5$, the coefficient $c_l(\Omega)$ is of 
order $10^{-2}$ [for higher multipoles, $c_l$ decays very fast,
scaling roughly as $(1-\Omega)^l$].

Since $H_F^2 > (8\pi G/3) \Delta V$, where $\Delta V$ was introduced in
Eq. (\ref{rnot}), we have
\begin{equation}
H_F^2 R_0^2 >24\pi G {S_1^2 \over \Delta V} \approx 16 G H_T^2
\left({c_l \over D_l^W}\right).\label{inequality}
\end{equation}
From (\ref{mugeneral}), and using
(\ref{deltas}), (\ref{deltasc}) and (\ref{inequality}),
we find
\begin{equation}
\mu \gtrsim \epsilon^2 K_l
            \left({D_{10}^S \over D_l^{SC}}\right)^2
            \left({D_{10}^S \over D_l^W}\right)^2 {1\over D_{10}^S}, 
\label{largemu}
\end{equation}
where the coefficient
$K_l \equiv {400 \pi^2}{c_l^2 d_l^2 b_l^2/ b^3_{10}}$ is plotted in Fig. 4
for various values of $l$ and $\Omega$.
The inequality (\ref{largemu}) turns out to be somewhat restrictive.

\begin{figure}[t]
\centering
\hspace*{-4mm}
\leavevmode\epsfysize=10cm \epsfbox{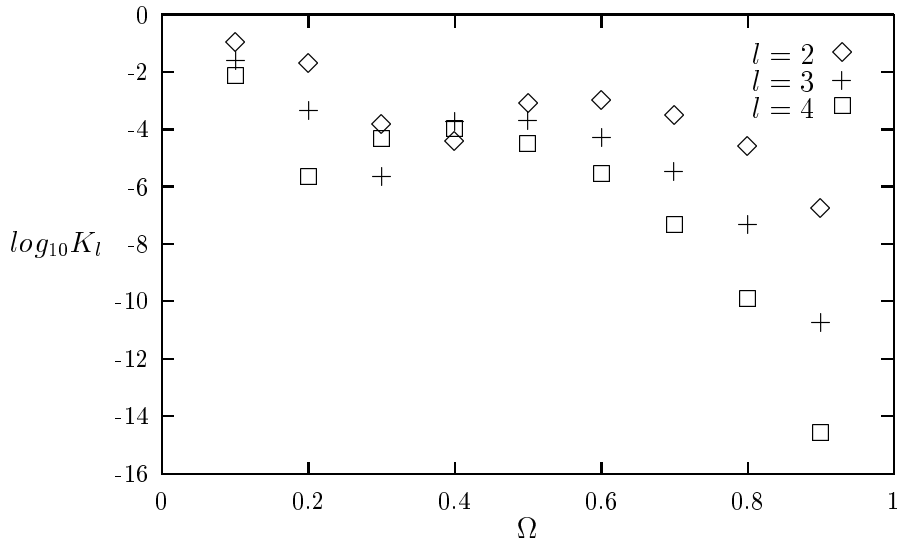}\\[3mm]
\caption[fig4]{\label{fig4} The coefficient $K_l$ for
various values of the density parameter}
\end{figure}

In the model given in Eq. (\ref{coupled}), 
the parameter $\epsilon=[2N(\phi_0)]^{-1}$ is of order $10^{-2}$.
From Fig. 4, the coefficient $K_l$ is never smaller than 
$10^{-4}$ for $\Omega$ in the range $.1$ to $.7$. 
Hence, we find
\begin{equation}
\mu \gtrsim 100 \left({D^S \over D^{SC}}\right)^2
            \left({D^S \over D^W}\right)^2 \left({10^{-10}\over D^S}\right).
\label{ineq}
\end{equation}
As discussed in Section IV, constraints from the observed fraction of clustered
matter imply $\mu \lesssim 3$. On the other hand, observations of CMB
anisotropies require $C^i_l < 10^{-10}$.
Hence, we conclude that this model can only
satisfy all observational constraints 
if CMB anisotropies are not completely
dominated by scalar subcurvature modes.  

If the observed CMB anisotropies are due to inflation, then we
should have $D^{SC},D^W\lesssim D^S\sim 10^{-10}$, and Eq.(\ref{ineq})
gives $\mu\gtrsim 100$.  For such values of $\mu$, the peak of the
probability distribution is at very low values of $\Omega$, and the
corresponding fraction of clustered matter is unacceptably small.  
It is therefore unlikely that
the two-field potential (\ref{coupled}) can give a realistic model of
open inflation which will explain both a nontrivial value of $\Omega$
and the observed spectrum of CMB fluctuations.
 
This problem disappears if the observed CMB anisotropies are due to a
different source, such as cosmic strings or other topological defects
forming at the end of inflation, which would also be responsible for 
structure formation. Also, the restriction (\ref{largemu}) will be less
severe if the observed value of $\Omega$ is larger than $.7$, since the 
coefficient $K_l$ is then much smaller, or in models with a smaller 
slow roll parameter $\epsilon\lesssim 10^{-3}$.

\section{The cosmic age coincidence}

The usual objection against models with $\Omega<1$ is that it is hard to 
explain why we happen to live at the epoch when the curvature is about 
to dominate. That is, why
$$
t_0 \sim t_c,
$$
where $t_0$ is the present time and $t_c$ is the time of curvature 
domination. Observers at $t\ll t_c$ would find $\Omega \approx 1$,
while observers at $t\gg t_c$ would find $\Omega\ll 1$. It appears that
one needs to be lucky to live at the time when $\Omega \lesssim 1$.
There is another coincidence which is required in open-universe models
and which also calls for an explanation. Observationally the epoch of 
structure formation, when giant galaxies were assembled, is at 
$z\sim 1 - 3$, or $t_G\sim t_0/3 -t_0/8$. On the other hand, the 
interesting range of $\Omega$ for open universe models is 
$0.3<\Omega < 0.9$, which corresponds to $z_c \sim 0.1- 2$, or
$t_c \sim 0.3 t_0-0.9 t_0$. We see that $t_G$ and $t_c$ are within one
order of magnitude of one another. It is not clear why these seemingly
unrelated times should be comparable. We could have for example
$t_G \ll t_c$. In this Section, we shall argue that the coincidence 
$$
t_G\sim t_c \sim t_0
$$
may be not as surprising as it first appears.

Let us begin with the coindidence $t_G\sim t_c$. 
In models we are considering here, most of the volume in each
quasi-open bubble is occupied by regions with small values of $\Omega$,
corresponding to small values of $t_c$.  Mathematically this is
expressed by the fact that the ``dehumanized'' probability
distribution $d{\tilde {\cal P}}(\Omega)$ in Eq.(14) is peaked at
$\Omega = 0$ (for $\mu>1/2$).  On the other hand, the ``human factor''
$\nu(\Omega)$ suppresses all values of $\Omega$ for which $t_c<t_G$,
so that curvature domination interferes with structure formation.  As
a result, the peak of the full probability distribution $d{\cal P}(\Omega)
= \nu(\Omega)d{\tilde {\cal P}}(\Omega)$ is shifted to a value of
$\Omega$ corresponding to $t_c\sim t_G$.  Hence, we should not be
surprised that $t_c\sim t_G$ in our universe.

It remains to be explained why we live at a time $t_0\sim t_G$.  Clearly,
$t_0$ could not be much less than $t_G$, so we need to explain why we
do not have $t_0\gg t_G$.  We now recall Dicke's observation
\cite{Dicke} that the
time $t_0$ is unlikely to be much greater than $t_G+t_{\star}$, where 
$t_{\star}\sim 10^{10}$ yrs is the lifetime of a typical main sequence
star.  Noticing that $t_{\star}\sim t_G$, we conclude that the expected
value of $t_0$ is $\sim t_G$.

The value of $t_{\star}$ and $t_G$ depend only on fundamental
constants and on the amplitude of the cosmological density
fluctuations.  In the models we have considered in this paper, where
$\Omega$ is the only variable parameter, these timescales are fixed
and one cannot address the question of why they are similar.

If cooling failure (discussed at the end of Section IV and in
Appendix C indeed
represents  a barrier for effective star
formation, then it adds 
yet another timescale which is comparable to the other four
we have encountered in this Section. This is the time $t_{cf}$ 
after which collapsing 
gas clouds of galactic mass cannot fragment and remain pressure
supported.  This 
timescale is also determined by fundamental constants, so the
coincidence of this scale with $t_G$ cannot be explained within 
our simple model. However, it is easy to generalize the model
so that both $\Omega$ and $\sigma_{rec}$ are variable.
For instance, instead of just one slow roll scalar field, we can consider 
two of them,
$$
V(\sigma,\phi_1,\phi_2)=V_t(\sigma)+{\sigma^2\over 2} 
(g_1 \phi_1^2 + g_2 \phi_2^2).
$$
In this case, the two slow roll fields will have different mass
inside the bubble. 
The duration of inflation and the amplitude of density perturbations
are determined by the point in the plane $(\phi_1,\phi_2)$
where the fields land after tunneling.
Changing to polar coordinates on that plane, the number of e-foldings of 
inflation depends basically on the radial coordinate $R$ (how far we 
are from the bottom of the potential). On the other hand, 
the amplitude of density perturbations depends on the effective mass 
along the curve described by the inflaton, which is determined by the 
angular coordinate $\Theta$. 

The volume factor in the probability distribution will be the 
same on $R=const.$ surfaces, whereas the tunneling factor will
choose the direction $\Theta$ in which the mass 
$m^2(\Theta)  \propto g_1 \cos^2\Theta + g_2 \sin^2\Theta$ 
is the lowest.  
In our model, $\sigma_{rec} \propto m(\Theta) N(R) M_P^{-1}$,
where $N(R) \sim G R^2$  is the number of e-foldings of inflation and 
$M_p$ is the Planck mass. Low $m$ means large $t_G$, because the 
smaller is $\sigma_{rec}$, the longer it takes for a perturbation
to go nonlinear. Hence, volume and tunneling factors would choose
the largest possible $t_G$. On the other hand, $t_G$ cannot be
larger than the cooling boundary $t_{cf}$. Therefore, $t_G \sim t_{cf}$
could also be explained in this model. 
This argument can be regarded as an explanation for the observed 
amplitude of density fluctuations $Q$ in our universe: the value $Q\sim 
10^{-5}$ is selected by the condition $t_G\sim t_{cf}$.\footnote{
Anthropic bounds on $Q$ have been previously discussed in Ref. 
\cite{TR}. Tegmark and Rees \cite{TR} used the inequality $t_G < t_{cf}$ 
to impose a lower bound on $Q$.  To obtain an upper bound, it has been 
argued \cite{AlexQ,TR} that for large values of $Q$ galaxies would be 
too dense and 
frequent stellar encounters would disrupt planetary orbits.  To estimate 
the rate of encounters, the relative stellar velocity was taken to be the 
virial velocity $v_{vir}\sim 200 km/s$, resulting in a bound 
$Q>10^{-4}$.  However, Silk \cite{silkp} has pointed out 
that the local velocity dispersion of stars in our galaxy is an order of 
magnitude smaler than $v_{vir}$.  This gives $Q > 10^{-3}$, which is a 
rather weak constraint.  This issue does not arise in the approach 
outlined in the text above, since in our case large values of $Q$ are 
suppressed by the tunneling and volume factors in the probability.}
A detailed analysis is left 
for further research.

\section{Summary and Conclusions}

We have calculated the probability distribution for the density parameter 
in models of open inflation with variable $\Omega$.
This probability is basically the product of three factors: the ``tunneling''
factor, which is related to the microphysics of bubble nucleation and
subsequent expansion; the volume factor, related to the amount of 
slow roll inflation
undergone in different regions of the universe; 
and the ``anthropic factor'', which determines the number of galaxies
that will develop per unit thermalized volume.  It is interesting that
the expression for the probability (\ref{distributiony}) depends on
the underlying particle 
physics model through a single dimensionless parameter $\mu$, defined
in Eq.(\ref{defmu}).

Taking the minimum of the slow roll
potential to be at $\phi=0$, the tunneling factor tends to suppress large 
initial values of $\phi$, favouring low values of $\Omega$.
However, only those regions for which $\phi$ is large enough will inflate.
Hence, there will be a competition between 
volume enhancement and ``tunneling'' suppression.

The most interesting situation occurs when the tunneling suppression dominates
over the volume factor. In this case, the product of both would peak
at $\Omega=0$, and the anthropic factor $\nu(\Omega)$ 
becomes essential in determining
the probability distribution. In an open universe, cosmological
perturbations stop growing when the universe becomes curvature
dominated, and for low values of $\Omega$ structure formation is
suppressed.  The effect of the anthropic factor is, therefore, to
shift the peak of the distribution from $\Omega=0$ to a nonzero value
of $\Omega$.

 
As a first approximation \cite{VW,MSW},
we have taken $\nu(\Omega)$
to be proportional to the fraction of matter that clusters on the 
galactic mass scale in the entire history of a certain region. 
We have found that the 
peak of the distribution is given by the condition
\begin{equation}
\kappa \left({1-\Omega \over \Omega}\right)_{peak} \approx 
\left({3\over 2}\mu-{5\over 4}\right)^{1/2},
\label{mon}
\end{equation}
where the coefficient $\kappa \sim 10^{-1}$ is defined in (\ref{kappa}).
For models with $\mu \sim 1$ (which can be easily constructed), 
the probablility distribution for the density parameter ${\cal P}(\Omega)$ 
can peak at values of $\Omega$ such that 
$x=(1-\Omega)/\Omega\sim 1$ (See Fig.
1). The peaks are not too sharp, with amplitude $\Delta y \approx 1/2$,
or $\Delta x \approx 5$, so a range of values of $\Omega$ would be measured
by typical observers.

The analysis we presented here demonstrates that, given a
particle physics model, the probability distribution for $\Omega$ can
be unambiguously calculated from first principles.  We can also invert
this approach and use
our results to exclude particle physics models which give the peak of
the distribution at unacceptably low values of $\Omega$.
This gives the constraint $\mu\lesssim 3$.

An independent constraint on the model parameters can be obtained from CMB
observations.  If the observed CMB anisotropies are to be explained
within the same two-field model of open inflation, without adding any
extra fields, then we have shown in Section IV that the corresponding 
constraint (if the observed value of $\Omega$ lies in the range $.1$ 
to $.7$) is
$\mu \gtrsim  10^{6} \epsilon^2$, where 
$\epsilon$ is the slow roll parameter defined in (\ref{epsilon}).
Combinig both constraints, we obtain a bound on the slow roll parameter 
$$
\epsilon\lesssim 10^{-3}.
$$
This bound is somewhat restrictive. For instance,
for the simple free field model (\ref{coupled}), the slow roll 
parameter is of order $10^{-2}$, and so this model would contradict
observations. It is easy, however, to generalize the slow roll 
potential in order to make $\epsilon$ sufficiently small. 
If one allows some other source for CMB fluctuations (e.g.,
topological defects), then the CMB constraint is much less
restrictive, and simple models of the form (\ref{coupled}) are still
viable.

We have advanced anthropic arguments towards explaining the
``cosmic age coincidence'', that is, whether it would be surprising
to find that we live at the time 
when the curvature is about to dominate. We have argued
that this is not unexpected. We have also discussed a three-field model 
in which the amplitude of 
density fluctuations $Q$ becomes a random variable.  We have outlined an 
argument explaining the observed value $Q\sim 10^{-5}$ in the framework 
of this model.

While this work was being completed, Hawking and Turok \cite{HT98},
have suggested the possibility of creation of an open universe from
nothing (see also \cite{everybody}). The validity of the
instantons describing this process \cite{alex}, and also their ability to
successfully reproduce a sufficiently homogeneous universe, is still a 
matter of debate and needs further investigation. Clearly, the analysis
presented in this paper can be easily adapted to this new framework.

\section*{Acknowledgements}

We would like to thank Andrew Liddle, Martin Rees and Max Tegmark 
for useful discussions.
J. G. thanks the Tufts Cosmology Institute for hospitality.
This work has been partially supported by NATO under grant 
CRG 951301. J.G. acknowledges support from CICYT under contract AEN95-0882.
A.V. Acknowledges support from the National Science Foundation.

\section*{Appendix A: Size of the populated universe and classical
anisotropies}

As mentioned in Section II, a quasi-open universe is formed by an ensemble 
of inflating regions of very large size compared to the curvature scale.
Clearly, the central parts of each region will inflate longer and, will
have a larger density parameter than the peripheric regions. Hence, the
fraction of clustered matter will decrease as we move away from the 
center. Here we shall estimate the size of the populated region, which,
as we shall see, is much smaller than the size of the inflating region.

From (\ref{peak}), (\ref{omegaphi}) and (\ref{efoldings}) we have
\begin{equation}
d\ln \nu|_{peak} = 
2\pi G(6\mu - 3) d \phi_0^2 \sim  
{\phi_0^2 \over f^2} d\ln\phi_0.
\label{nuphi}
\end{equation}
This equation gives the variation of $\nu$ due to the gradients in $\phi$ as 
we move away from a typical civilization which measures the peak value of
$\Omega$. [the estimate in (\ref{nuphi}) holds provided that $\mu$ is not 
too close to $1/2$, say $\mu \geq .6$].

Taking this civilization to be located at $r=0$, the
gradients can be decomposed in multipoles. 
For $l=0$, $d\phi_0$ can be found from 
(\ref{island}). For $r\sim 1$ (which for low $\Omega$ 
roughly corresponds to the present Hubble distance) we have 
$d \ln\phi_0 \sim \gamma$. Combining with
(\ref{nuphi}) we find that $\nu$ changes by 
\begin{equation}
\delta \ln\nu \sim {\gamma (\phi^{peak}_{0})^2 \over f^2} \equiv X
\label{x}
\end{equation}
over the Hubble distance. 

For $X\ll 1$, $\nu$ would
not change appreciably on cosmological scales. Using (\ref{island}),
the co-moving size 
of the populated universe can be estimated as the distance at which
$\nu$ drops by an order of magnitude,
$$r_p\sim X^{-1}.$$ 
For $\mu>{1/2}$ we need $\phi_0^2\gg f^2$ in order to have 
sufficiently long inflation. Hence 
we find that the size of the populated region
is larger than the curvature scale but still
much smaller than the size of the quasi-open island,
$1\ll r_p\ll \gamma^{-1}$.

For $X\gg 1$ we can use (\ref{island}) for small $r$ to obtain
\begin{equation}
d\ln\phi_0 \approx \gamma {r^2 \over 6}
\end{equation}
In that case, the size of the populated universe can be estimated as
\begin{equation}
r_p\sim X^{-1/2}\ll 1,
\end{equation} 
and the human factor would drop by several orders of magnitude
within our Hubble radius. Clearly, we should not expect to
lie precisely at the center of the hospitable region, but rather 
at the outskirts, and then we would observe a large anisotropy in
$\nu$ around us. This can be confirmed by analyzing the
$l>0$ supercurvature modes. The amplitude of $l>0$ 
modes is of order \cite{GGM}
\begin{equation}
\delta\phi_0\sim \gamma^{1/2} f r^l.
\label{higher}
\end{equation} 
Combining with (\ref{nuphi}) we have
$$
\delta\ln \nu \sim X^{1/2} r^l.
$$
For large $X$, the $l=1$ anisotropy in $\nu$ becomes of order one
at the distance $r_p$, as expected. 

If $\nu$ is proportional to the fraction of clustered matter, as we 
have assumed in the preceding section, a large drop
in this quantity is already excluded by observations \cite{obs}, 
so the constraint
\begin{equation}
X\equiv {\gamma (\phi_{0}^{peak})^2 \over f^2} \lesssim 1
\label{constraint1}
\end{equation}
must be imposed on our model. 

This constraint is relevant to the question of
classical anisotropies in a quasi-open universe, discussed in
Ref. \cite{GM}. To an observer living at large 
distances from the center of the island $r\gg 1$
the universe would look anisotropic, with
$d\phi_0 \sim \gamma \phi_0$ over the 
curvature scale around that point. For $X>1$ this anisotropy 
would be larger than the $l>0$ quantum fluctuations from supercurvature
modes (\ref{higher}). However, as shown above, for $X>1$
the typical observer must be at a distance 
$r \sim X^{-1} \lesssim 1$ from the center of the island, and the
arguments of Ref. \cite{GM} do not apply.  Hence, even though the
constraint (\ref{constraint1}) coincides with the one derived in
\cite{GM} (where a single island was considered and the universe 
was not taken to be homogeneous on
very large scales), its interpretation is very different.
It does not arise from requiring that the classical CMB anisotropy
should be smaller than the $l>1$ supercurvature anisotropy but from
demanding that the factor $\nu$ determining the density of civilizations 
should be isotropic around us.

On the other hand, for the simple model (\ref{coupled}),
one can find a much stronger constraint on $X$ by 
combining the bounds from the observed isotropy of the CMB discussed 
in Section VI, with the bounds on the observed fraction 
of clustered matter. Indeed, the supercurvature anisotropy 
can be expressed as
\begin{equation}
D_l^{SC}\sim 10^{-6} \mu^{-2} X,\label{deltas2}
\end{equation}
where $X$ was defined in (\ref{x}).
Using the constraints form the observed fraction of clustered matter 
$\mu \lesssim 3$ [see(\ref{rangemu})] and requiring 
$D_l^{SC}\lesssim 10^{-10}$, 
this results in
\begin{equation}
X\lesssim 10^{-4} \mu^2 \lesssim 10^{-3},
\label{constraint2}
\end{equation}
a much stronger constraint than (\ref{constraint1}). Hence, the 
size of the populated universe should be at least $10^3$ times
larger than the curvature scale in this model.

\section*{Appendix B: Evaluation of $\kappa$}

As mentioned at the end of Section VI, 
in order predict the expected values of $\Omega$ at $T_{CMB}=2.7 K$ in our part
of the universe, we need to know $\mu$, as well as the coefficient
$$
\kappa = {3\over 5\sqrt{2} \sigma_{rec}}
{T_{CMB}\over T_{rec}}
$$  
that relates $x$ to $y$. For the temperatures we take $T_{CMB}\approx 2.7 K$
and $T_{rec}= 1100\, T_{CMB}$. The main unknown in this coefficient is 
$\sigma_{rec}$. 

The value of $\sigma_{rec}$ is, to a very good approximation, 
the same in all thermalized regions. Hence it can be inferred from
measurements of CMB anisotropies on large angular scales in our
observable region. Since we are interested in relatively small
scales, we also need to make some assumptions about the power spectrum of 
density fluctuations. We shall take a scale invariant
cold dark matter (CDM) adiabatic spectrum. 
As we shall see, our ability to infer the
precise value of $\sigma_{rec}$ will be  limited by the fact that the
density parameter $\Omega_0$ in the observable part of our 
universe is not known very 
precisely. Hence, we shall leave it as a free parameter.
We emphasize that $\Omega_0$ is the
value of the density which is actually realized in our universe today, and
whose precise value we do not know yet. This should not be confused
with the random variable $\Omega$ which appears in the probability 
distributions, and which takes different values in different regions.

In order to determine $\sigma_{rec}$, we note that
\begin{equation}
\sigma_{rec}(R) = A^{-1}(\Omega_0) \sigma_0(R),
\label{gal}
\end{equation}
where $\sigma_0(R)$ is the present density contrast on the relevant scale $R$
and $A(\Omega_0)$ is the factor by which linear perturbations 
have grown from the time of recombination until the present time.
In an open universe, this factor is given 
by \cite{peebles}
$$
A(\Omega_0) = {5\over 2} {f(x_0) \over x_{rec}},
$$
where $x_0$ and $x_{rec}$ are the values of $(1-\Omega)/\Omega$ 
in our observable universe at present and at recombination respectively. 
The function $f$ is given in (\ref{function}).

With this, we have
$$
\kappa = {3 \over 2 \sqrt{2} \sigma_{0}}{f(x_0) \over x_0},
$$
where we have used the fact that $xT=const.$ in the matter era.

The present linear density contrast $\sigma_0$ is given by 
\cite{lid95,MSW}
\begin{equation}
\sigma_0(R_{gal})  = (c_{100} \Gamma)^2 \delta_H K^{1/2}(R_{gal}).
\label{sigmanot}
\end{equation}
Here $c_{100} \approx 2997.9$ is the speed of light in units of
$100 km s^{-1}$, $\delta_H$ is the dimensionless amplitude at horizon
crossing (which can be inferred from COBE measurements), 
$\Gamma = \Omega_0 h$ is the ``shape parameter'', 
with $h$ the present hubble rate in units of $100 km s^{-1} Mpc^{-1}$ 
(we ignore the effect of baryon density in this expression for $\Gamma$),
and $K$ contains the information on the power spectrum and the length scale
$R_{gal}$ we are considering. 

For a scale invariant spectrum, $K$ is given by \cite{MSW}
$$
K(R) \equiv \int_0^{\infty} q^3 T^2(q) W^2(q R h \Gamma Mpc^{-1})\ dq.
$$
where the transfer function $T$ in the CDM model 
can be approximated as \cite{bardeen}
$$
T(q)={\ln (1+2.34 q)\over 2.34 q} [1+3.98 q+(16.1 q)^2 +
(5.46 q)^3+(6.71 q)^4]^{-1/4},
$$
and the top-hat window function $W$ in momentum space is given by
$$
W(u) = {3\over u^3}(\sin u - u \cos u).
$$
In order to find numerical estimates, we shall 
consider \cite{MSW} $R_{gal}= 1-2\ h^{-1} Mpc$. Roughly speaking, 
this corresponds to the scale whose baryon content collapses to form a 
galaxy with a mass comparable to that of the Milky Way. 
Also, by requiring that CDM predictions correctly reproduce the statistics 
for galaxy distribution on scales of tens of megaparsecs \cite{PD94},
the shape parameter is constrained to be in the range 
$$\Gamma \approx 0.25 \pm 0.05\ .$$
For our estimates, we shall take $\Gamma = .25$. With this, we find
$$
K(1\ h^{-1} Mpc) \approx 0.049, \quad K(2 h^{-1} Mpc) \approx 0.026\ .
$$ 

For the dimensionless amplitude $\delta_H$ we shall use the fitting
function given by Liddle et al. \cite{lid95}
\begin{equation}
\delta_H(\Omega) = (4.10 + 8.83 \Omega - 8.50 \Omega^2)^{1/2}\times 10^{-5}
\label{fit}
\end{equation}
Hence, the coefficient 
\begin{equation}
\kappa={3 \over 2 \sqrt{2} (c_{100} \Gamma)^2 K^{1/2}} 
{f(x_0) \over \delta_H(\Omega_0) x_0}
\end{equation}
will be sensitive to our ignorance of the value of $\Omega_0$ in 
our universe, as mentioned above. 

In Fig. 1 we plot $\kappa$ as a function of $\Omega_0$ for the two
chosen values of the scale $R_{gal}$.

\section*{Appedix C: Effects of cooling failure}

As mentioned at the end of Section IV, fragmentation of gas clouds
will only occur if the 
cooling timescale $\tau_{cool}$ is smaller than the timescale 
needed for gravitational collapse $\tau_{grav}$. Because of
this, fragmentation will be suppressed after a certain critical 
time $t_*$.  Here we shall
investigate the possibility \cite{TR} that clouds collapsing at
$t>t_*$ do not effectively form stars even after they eventually
cool. We shall see that, as a consequence, the peak of the distribution will 
be shifted to somewhat larger values of $\Omega$. 

The density of the virialized collapsing cloud $\rho_{vir}$ is
given by \cite{TR,laco93}
$$
\rho_{vir} \sim 10^2 (G t_{vir}^2)^{-1},
$$
where $t_{vir}$ is the time at which the collapse occurs.
The virialization temperature can be
estimated as 
$T_{vir} \sim m_p v_{vir}^2 \sim m_p (G^3 \rho_{vir} M_{g}^2)^{1/3}$.
Here $m_p$ is the proton mass, and $v_{vir}$ is the virial velocity
$v_{vir}\sim (GM_{g}/L)^{1/2}$, where $L$ is the size of the collapsed object.
The later an object collapses, the colder and rarer it will be.

The cooling rate $\tau_{cool}^{-1}$ of a gas cloud of fixed mass 
depends only on its density and temperature, 
but as shown above both of these quantities are determined by 
$t_{vir}$ \footnote{Actually, the fraction
of baryonic matter $X_b$ is also relevant for cooling.
Following \cite{TR} we shall take $X_b \sim 0.1$}.
The timescale needed for gravitational collapse is $\tau_{grav}\sim t_{vir}$ 
Therefore, the condition $\tau_{cool}<\tau_{grav}$ gives an upper 
bound $t_*$ on the time at which collapse occurs. Matter that clusters after 
that time should not contribute to the anthropic 
factor $\nu(\Omega)$.

Various cooling processes
such as Bremsstrahlung and line cooling in neutral Hydrogen and Helium
were considered in Ref. \cite{TR}. 
For a cloud of mass $M_g \approx 10^{12} M_{\odot}$, 
cooling turns out to be  efficient 
\footnote{This upper bound on $t$ is determined by line cooling in Helium.
For $M_g \approx 10^{12} M_{\odot}$ there is also a narrow range of time 
near $t \approx 3 \cdot 10^{11} {\rm Yr}.$ 
where cooling is again efficient 
due to Hydrogen line cooling. However, the range is very narrow
and we shall disregard the galaxies which may form during this short 
late period.}
for 
\begin{equation}
t<t_* \approx 3 \cdot 10^{10} {\rm Yr}.
\label{rhoc}
\end{equation}
This value of $t_*$ should be taken only as indicative, since the
present status of the theory does not allow for very precise
estimates.


From the time of recombination to the
time $t_*$ fluctuations will grow by the factor \cite{peebles}
$$
G_*(\Omega) = {5\over 2 x_{rec}} f(x_*),
$$
whereas the critical density contrast is given by
$$
\delta_c = {3\over 2} f(x_*) g(x_*),
$$
where $f(x)$ and $g(x)$ are given in (\ref{function}) and (\ref{g}).
Following the steps that lead to Eq. (\ref{erfc1}) we now find
\begin{equation}
\nu={\rm erfc} \left[\kappa x g(x_*)\right].
\label{erfc2}
\end{equation}
Noting that in the matter era \cite{peebles}
$$
t= t_{rec}{x^{1/2}(1+x)^{1/2}- \sinh^{-1} x^{1/2}
\over x_{rec}^{1/2}(1+x_{rec})^{1/2}- \sinh^{-1} x_{rec}^{1/2}  }
$$
and using $x_{rec} \ll 1$ we have
$$
g(x_*)\approx 1+ \left({3\pi t_{rec} \over 2 t_*}\right)^{2/3}{1\over x_{rec}}
=1+{\Delta_* \over x}
$$
where 
\begin{equation}
\Delta_* = {T_{rec}\over T_{CMB}}\left({3\pi t_{rec} \over 2 t_*}\right)^{2/3}.
\end{equation}
Therefore, the fraction of matter that clusters on a given scale before 
the critical time is basically obtained by shifting $y$ in Eq. (\ref{erfc1})
by the constant $\kappa\Delta_*$
\begin{equation}
\nu={\rm erfc} (y+\kappa\Delta_*)
\end{equation}
Using the values $\Omega_0=.5$ and $h=.5$ for our observable universe 
in order to infer $\kappa$ (see Fig. 1) and $t_{rec}\approx 5.6\cdot 10^{12} 
(\Omega_0 h^2)^{-1/2}{\rm s}$, we have $\kappa \Delta_* 
\approx 0.2$ (as in Appendix B, we have used $T_{rec}= 1100\, T_{CMB}$).
The peak of the modified probability distribution is plotted in 
Fig. 3 (curve $c$) as a function of $\mu$, next to the original 
curve $a$ where cooling failure is neglected. 
Asymptotically both curves differ only by 
$\Delta y_{peak}\approx \kappa \Delta_*/2=0.1$.
This is much smaller than the width of the distribution $\Delta y
\sim 0.5$, so the effect is rather small.
In Fig. 5 $(a)$ we plot the probability distribution as a function of
$x=(1-\Omega)/\Omega$ (at the temperature 
$T_{CMB}=2.7 K$) for three different values of $\mu$, without taking into 
account cooling effects. For comparison, in Fig. 5 $(b)$ we show the 
modified distribution when matter that clumps after time $t_*$ is disregarded.
 
We note that even if cooling is efficient, the density of the protogalactic
cloud is likely to affect the number and the mass distribution of stars in
the resulting galaxy. Masses of suitable stars should be large enough to
provide the necessary luminosity and small enough so that the stellar
lifetime is sufficient to evolve intelligent life. It is conceivable that
the number of such stars drops with the density, in which case 
the upper bound on $t$ should be stronger than (\ref{rhoc}).
Again, galaxy formation is not understood to the extent that would allow 
us to estimate this upper bound with accuracy. However,
since we do not observe many giant galaxies forming at redshifts 
lower than $z=2$, we may consider as a third possibility 
the case where matter that clumps after the time 
$t_* \approx 3 \cdot 10^9 {\rm Yr}$ is excluded from the anthropic factor. 
This corresponds to $\kappa\Delta_* \approx 1$.  
Even in this extreme case the shift in the peak 
$|\Delta y_{peak}|\leq \kappa \Delta_*/2 \approx 0.5$ is of the
same order of magnitude as the width of the distribution $\Delta y \sim 0.5$ 
[see (\ref{deltaomega})]. The new distribution as a 
function of $x$ is plotted in Fig. 5 $(c)$.

Therefore, we find that the impact of these effects on the 
probability distribution may be significant, but not dramatic, and 
(\ref{omegapeak}) is still valid by order of magnitude.

\vfill\break

\begin{figure}[t]
\centering
\hspace*{-4mm}
\leavevmode\epsfysize=12cm \epsfbox{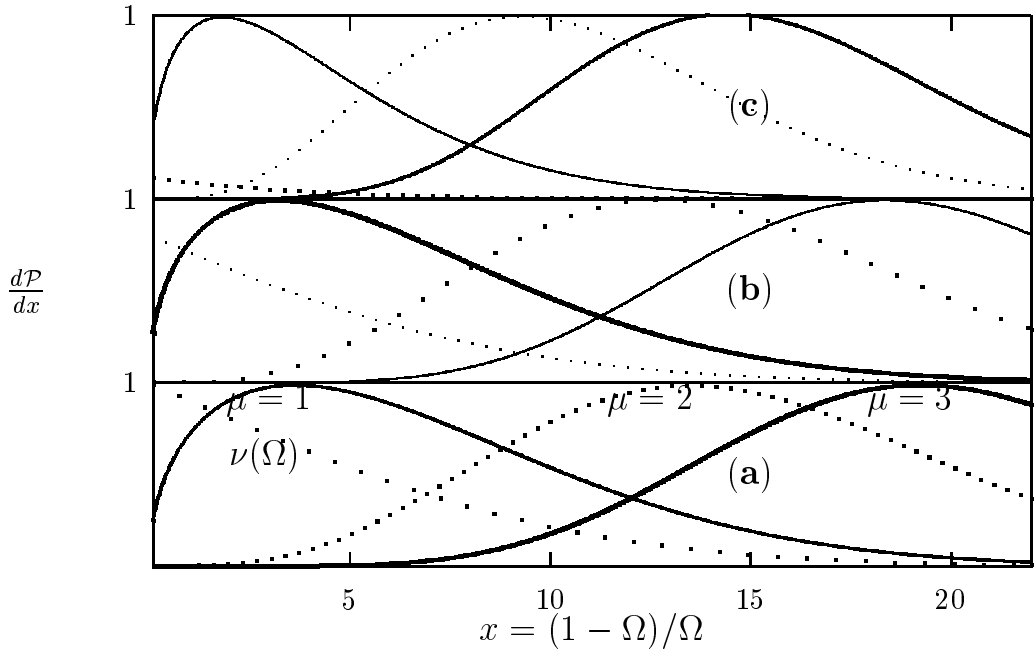}\\[3mm]
\caption[fig5]{\label{fig5} The probability distribution for $\Omega$
is sensitive to the fact that objects which collapse at very late times
have very low density, and therefore may be unsuitable for life.
Neglecting these ``selection'' effects,
frame (a) shows the probability distribution for $\Omega$ for various
values of $\mu$ (The value of $\Omega$ is the one measured at the 
temperature $T_{CMB}=2.7 K$). 
In this case, the anthropic factor
$\nu(\Omega)$ (also shown in the plot) is just proportional to the total 
fraction of matter that
clusters on the galactic mass scale in the entire history of a particular
region. In frame (b) we disregard matter which clusters after the time
when helium line cooling becomes inefficient, so that the collapsed
galactic mass objects cannot fragment into stars. Finally, as a more
extreme case, in frame (c) we disregard matter that clumps after the time 
$t_*\approx 3 \cdot 10^{9} {\rm Yr}$, since we do not see many giant 
galaxies forming at redshifts lower than $z=2$. }
\end{figure}


\begin{thebibliography}{999}
\bibitem{gott} A very similar approach was used by Carter \cite{carter},
Leslie \cite{leslie} and Gott \cite{rgott} to estimate the expected 
lifetime of our species. Gott also applied it to estimate the 
lifetimes of various political and economic structures, including the 
journal ``Nature'' where his article was published.
\end{thebibliography}
\end{document}